\documentclass[letterpaper]{article} 
\usepackage{aaai22}  
\usepackage{times}  
\usepackage{helvet}  
\usepackage{courier}  
\usepackage[hyphens]{url}  
\usepackage{graphicx} 
\usepackage{natbib}  
\usepackage{caption} 
\usepackage{amsthm,amsmath,amssymb} 
\usepackage{algorithm}
\usepackage{graphicx}
\usepackage{algpseudocode}
\usepackage{algorithmicx}
\usepackage{colortbl}  
\usepackage{tabularx}
\usepackage[table,xcdraw]{xcolor}
\usepackage{booktabs}
\usepackage{subcaption}
\usepackage{multirow}
\usepackage[normalem]{ulem}
\usepackage{float}

\newcommand{\DKL}{D_\text{KL}}
\newcommand{\DED}{D_\text{ED}}
\usepackage{times}
\usepackage{helvet}
\usepackage{courier}
\usepackage{newfloat}
\usepackage{listings}
\DeclareCaptionStyle{ruled}{labelfont=normalfont,labelsep=colon,strut=off} 
\floatstyle{ruled}
\newfloat{listing}{tb}{lst}{}
\floatname{listing}{Listing}
\title{Leveraging Modality-specific Representations for Audio-visual Speech Recognition via Reinforcement Learning}
\author{
    Chen Chen\textsuperscript{\rm 1}, Yuchen Hu\textsuperscript{\rm 1}, Qiang Zhang\textsuperscript{\rm 2, 3}, Heqing Zou\textsuperscript{\rm 1}, Beier Zhu\textsuperscript{\rm 1}, and Eng Siong Chng\textsuperscript{\rm 1}
    \\
}
\affiliations{
    \textsuperscript{\rm 1}School of Computer Science and Engineering, Nanyang Technological University\\
    \textsuperscript{\rm 2}ZJU-Hangzhou Global Scientific and Technological Innovation Center, Hangzhou, China\\
    
    \textsuperscript{\rm 3}College of Computer Science and Technology, Zhejiang University\\

}

\begin{document}

\maketitle

\begin{abstract}
Audio-visual speech recognition (AVSR) has gained remarkable success for ameliorating the noise-robustness of speech recognition. Mainstream methods focus on fusing audio and visual inputs to obtain modality-invariant representations. However, such representations are prone to over-reliance on audio modality as it is much easier to recognize than video modality in clean conditions. As a result, the AVSR model underestimates the importance of visual stream in face of noise corruption. To this end, we leverage visual modality-specific representations to provide stable complementary information for the AVSR task. Specifically, we propose a reinforcement learning (RL) based framework called MSRL, where the agent dynamically harmonizes modality-invariant and modality-specific representations in the auto-regressive decoding process. We customize a reward function directly related to task-specific metrics (\emph{i.e.}, word error rate), which encourages the MSRL to effectively explore the optimal integration strategy. Experimental results on the LRS3 dataset show that the proposed method achieves state-of-the-art in both clean and various noisy conditions. Furthermore, we demonstrate the better generality of MSRL system than other baselines when test set contains unseen noises.    
\end{abstract}

\section{Introduction}

\noindent Background noise is inevitable in real world that can dramatically degrade the speech quality and intelligibility, thereby increasing the difficulty of speech recognition task~\cite{hu2022dual, hu2022interactive}. In noisy scenarios, human will unconsciously observe the mouth region of speakers, as such noise-invariant visual cues can provide useful information for the corrupted speech understanding~\cite{ma2021end}.\par 
Similar to this, the audio-visual speech recognition (AVSR) technique couples the audio and visual modalities, which has attracted increasing research interest for several years~\cite{noda2015audio}. Recent machine learning based AVSR methods successfully demonstrate that deep neural network (DNN) can process and fuse raw acoustic and visual inputs to improve the noise-robustness for recognition through a supervised learning paradigm~\cite{petridis2018audio,zhou2019modality}. Additionally, self-supervised representation learning has been explored to capture the correlations between audio and visual lip movements for AVSR task, which has brought remarkable performance gain in terms of word error rate (WER) metric~\cite{shi2022learning}.\par

\begin{figure}[t]
  \centering
  \includegraphics[width=0.42\textwidth]{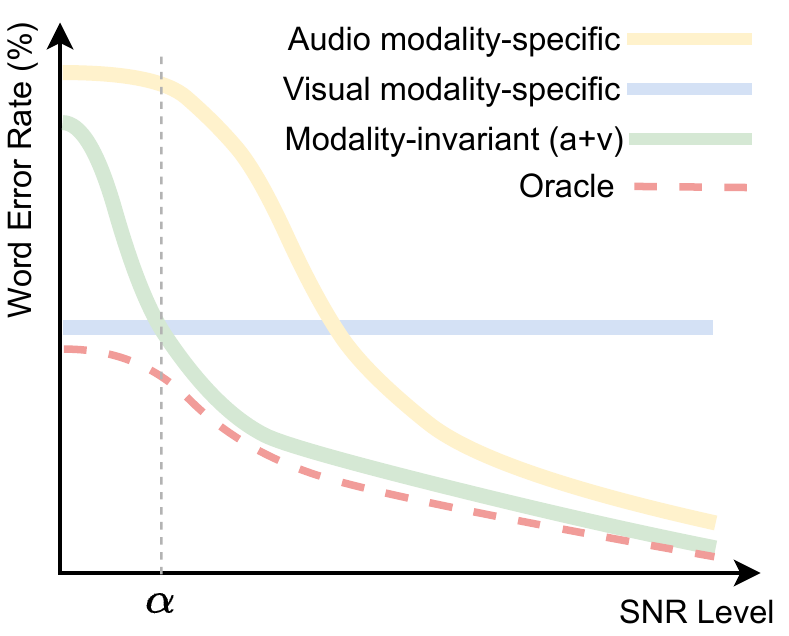}
  \caption{Research problem. ``SNR'' denotes the signal-to-noise ratio, and $\alpha$ is the threshold that modality-invariant representations lose the effectiveness compared with visual modality-specific representations. }
  \label{fig1}
\end{figure}

Mainstream AVSR methods focus on learning modality-invariant representations by fusing audio and visual modalities into a common subspace~\cite{song2022multimodal}.
However, such a fusion manner is prone to over-reliance on the audio modality, as it is much easier to recognize than video stream in clean conditions \cite{mittal2020m3er}. With the rise of noise levels, the importance of the video stream is increasingly underestimated in AVSR systems, and leads to sub-optimal performance since audio modality has already been corrupted by noise. We further illustrate this research problem in Fig~\ref{fig1}. Though modality-invariant representations (green line) outperform audio modality-specific representations (yellow line) by a large margin, it is still vulnerable to low signal-to-noise ratio (SNR) conditions. It is worth noting that when SNR is lower than $\alpha$, the modality-invariant representations even perform worse than visual modality-specific representations (blue line) which are completely unaffected by noise. We argue that this problematic situation can be avoided by reasonable coordination of modality-invariant and modality-specific representations, which is shown as the oracle system (red dashed line). \par
Although the significance of modality-specific representations has been emphasized in other multi-modal tasks, such as emotion recognition~\cite{hazarika2020misa}, it still remains challenging to integrate them into AVSR system for several reasons. Firstly, the real-world noises have dynamic and non-stationary temporal distributions, which confuse the recognizer to estimate the importance of visual modality-specific representations during auto-regressive decoding. Secondly, due to the natural distinction of input modalities, a uniform training schedule probably results in the vanishing gradient when we add a further sub-network to extract visual modality-specific representations~\cite{yao2022modality}. Finally, with parameter growth of neural networks, the existing integration strategies for new representations are prone to over-fit to specific types of noise distribution~\cite{fu2022reinforcement}, thereby failing to adapt to unseen noises in the wild. \par 
In this work, we aim to improve the AVSR system by leveraging visual modality-specific representations that carry noise-invariant information from the visual stream. To this end, we employ a pre-trained vision model that takes lip movement information as input and generates independent probability distribution for sequence generation. This idea is inspired by the language model rescoring that has been widely applied in popular ASR methods~\cite{song2022language,xu22i_interspeech,chen2022noise}. However, instead of applying a typical integration approach, \emph{e.g.}, late fusion~\cite{inaguma2019transfer}, we propose a reinforcement learning (RL) based method to dynamically harmonize the integration process in terms of the task-specific metric. RL is appropriate to play this integration role for: 1) The auto-regressive decoding of AVSR can be modeled as an RL formulation~\cite{bahdanau2016actor}, where the agent can consider multiple information for reasonable token prediction. 2) The customized reward function of RL bridges the training criterion and WER, thus encouraging it to effectively improve the model performance. 3) The beam search of inference step can provide a set of hypotheses for RL sampling, which allows the agent to explore the optimal policy on token level~\cite{chen2022self}. \par
The main contributions of this paper can be summarized as following:
\begin{itemize}
    \item We propose MSRL -- a novel AVSR system that utilizes visual modality-specific representations to dynamically remedy the noise-corrupted audio modality.
    \item MSRL adopts an RL-based integration method, where a new reward function is designed to encourage the agent to efficiently explore the optimal strategy in terms of the task-specific metric.
    \item Experimental results on the largest public LRS3 dataset show that MSRL is effective and achieves state-of-the-art performance in both clean and various noisy conditions. Furthermore, the comparative experiments on unseen noises demonstrate that MSRL has better generalization and adaptability than a strong baseline.
\end{itemize}\par

\section{Related work}
\noindent\textbf{Audio-visual speech recognition.}
Recently, AVSR has been attracting increasing research interest as it provides a potential solution for noise-robust speech recognition. To process and fuse audio-video modalities, TM-seq2seq~\cite{afouras2018conversation} applies a separated Transformer encoder for two modalities and fuses them before decoding.~\cite{ma2021end} presents a hybrid CTC/Attention model based on Resnet-18 and Conformer~\cite{gulati2020conformer}, which can be trained in an end-to-end manner.~\cite{tao2018aligning} demonstrate the importance of aligning two modalities before fusing them. Moreover, the AV-HuBERT~\cite{shi2022learning} learn the correspondence of audio and video modalities in a self-supervised manner, which is further augmented in~\cite{shi2022robust} to improve noise-robustness. 

\noindent\textbf{Modality-invariant and modality-specific representations.} Despite the advanced fusion techniques in multi-modal tasks~\cite{chen2022interactive}, prior works suggest that the model can benefit from modality-specific representations which capture some desirable properties~\cite{xiong2020modality}. Nevertheless, how to effectively utilize it is still an open question to be explored. MISA~\cite{hazarika2020misa} maps the multi-modal inputs into two subspaces for modality-invariant and modality-specific representations and then fuses them for final classification. MuSE~\cite{yao2022modality} employs separated encoders for multiple modalities, then harmonize them using different learning rates and late-fusion. Similarly,~\cite{feng2019learning} constructs an individual network for each modality, as well as designing a modality-shared identity loss to facilitate the extraction of modality-invariant representation. The integration of modality-specific representations is particularly difficult in an AVSR system because the single decoder is hard to dynamically weight the representations in a sequential decision process.\par

\noindent\textbf{Reinforcement learning in sequence generation.} Extensive existing works have demonstrated that RL is suitable to play an optimizing role in sequence generation tasks. In captioning tasks like image captioning~\cite{rennie2017self} and audio captioning~\cite{mei2021encoder}, a self-critical training approach based on RL can optimize the trained model in terms of non-differentiable metrics. Such an idea is also expanded to ASR tasks with a customized reward function~\cite{tjandra2019end, chen2022self}. Additionally, actor-critic based RL optimization algorithms are designed to improve task-specific score (\emph{e.g.} BLEU) in machine translation task~\cite{williams1992simple, bahdanau2016actor}. Compared with previous work, the proposed MSRL commits to the stability of RL training, where 1) we utilize pre-trained models to provide learned representations as state space, and 2) we design a reward function to encourage the policy network to explore in trust region~\cite{schulman2015trust}.

\begin{figure*}[t!]
  \centering
  \includegraphics[width=0.90\textwidth]{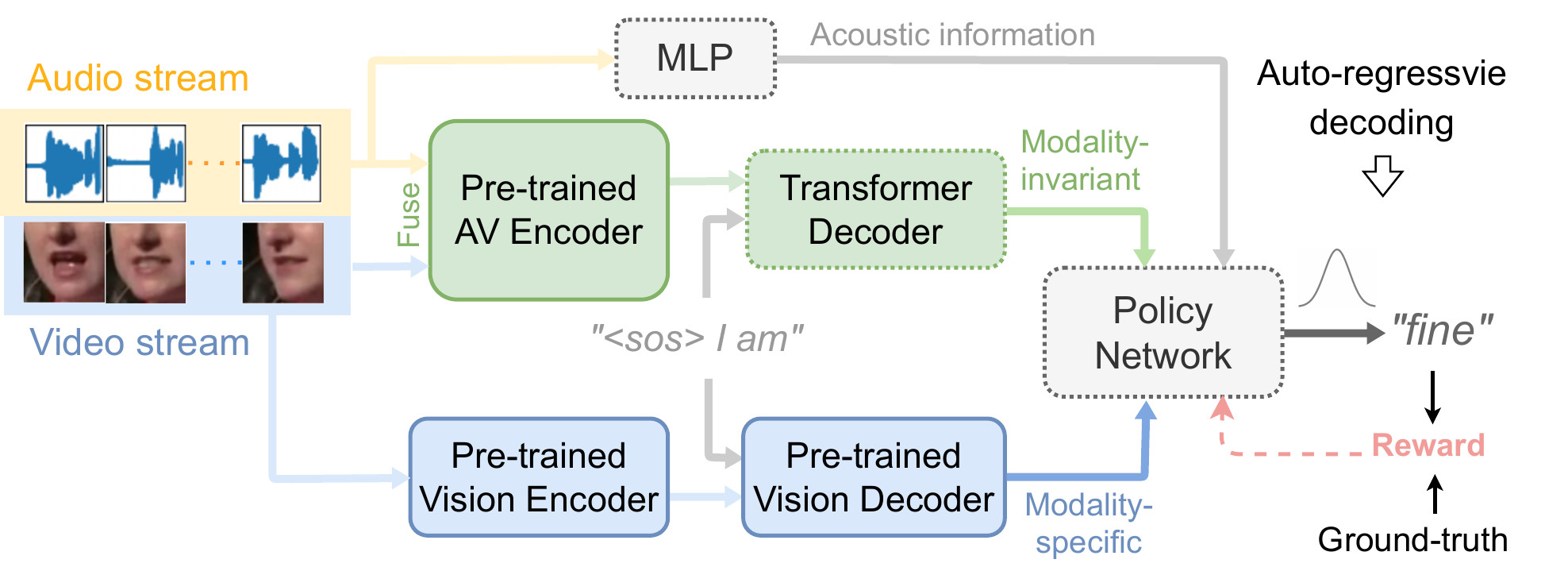}
  \caption{The block diagram of the proposed MSRL system. The solid box denotes such module is fixed during training, while the dashed box denotes it is trainable. The red dashed arrow denotes the process of back-propagation. Policy network considers multiple information in auto-regressive decoding and predicts the current token ``fine".}
  \label{fig2}
  \vspace{-0.3cm}
\end{figure*}

\section{Methodology}
In this part, we first introduce the main structure of the proposed MSRL system in Section~\ref{3-1}. Then in Section~\ref{3-2}, we model the auto-regressive decoding of AVSR as an RL formulation, as well as give mathematical derivation for the optimization. Finally, the training schedule of MSRL is illustrated in Section~\ref{3-3}.  

\subsection{Main Structure}
\label{3-1}
Given the acoustic utterance $A$ and its paired $l$-frame video stream $V=(v_1, v_2, ... ,v_l)$, the neural network of AVSR intends to predict a hypothesis sequence $Y=(y_1, y_2, ... , y_T)$. As shown in Fig.~\ref{fig2}, the audio and video streams are fed into a pre-trained AV encoder to extract hidden representation, where a ResNet block~\cite{he2016deep} and a linear layer are respectively served as front-end to obtain the audio and visual features. After concatenating them, a Transformer encoder~\cite{vaswani2017attention} with self-attention mechanism is employed to extract hidden representations. Subsequently, we utilize a learnable Transformer decoder including cross-attention mechanism to produce modality-invariant representations $F_i$ corresponding to the probability distribution of predicted tokens. Meanwhile, we further utilize a pre-trained vision model which similarly consists of ResNet front-end, Transformer encoder, and Transformer decoder. It consumes the video stream as input and independently produces visual modality-specific representations $F_v$ with the same shape as $F_i$. To harmonize the modality-specific and modality-invariant representations, a linear layer-based policy network are designed in the auto-regressive decoding process. Besides the $F_i$ and $F_v$ themselves, we argue that policy networks should also be aware of audio quality which is useful to estimate the importance of representations. To this end, an MLP block with 2 linear layers is used for downsampling and provides acoustic information $I_a$ for policy network. Finally, a combined distribution is generated to predict the current token (``fine" in Fig.~\ref{fig2}). \par
It is noted that all pre-trained models are fixed in the whole training process. The pre-trained AV encoder is initialized from AV-HuBERT~\cite{shi2022robust}, which captures cross-modal correlations between audio and video modalities by a self-supervised approach. The vision model~\cite{shi2022learning} including encoder and decoder is pre-trained via lip-reading task on LRS3 dataset, where only the video stream is required to generate target sequence. 

\subsection{RL Policy}
\label{3-2}
Basic Reinforcement Learning is typically formulated as a Markov Decision Process (MDP) that includes a tuple of trajectories $\left \langle \mathcal{S},\mathcal{A},\mathcal{R}, \mathcal{T} \right \rangle$. For each time step $t$, the agent consider state $s_t \in S$ to generate an action $a_t \in \mathcal{A}$ which interacts with environment. The transition dynamics $\mathcal{T}(s_{t+1}|s_t,a_t)$ is defined as transition probability from current state $s_t$ to next state $s_{t+1}$, and gain an instant reward $r_t(s_t,a_t)$. The objective of RL is to learn optimal policy to maximize the cumulative reward $\mathcal{R}$:
\begin{equation}
\mathcal{R} =  \mathop{\max}_{a_t\in \mathcal{A}} \sum_{t=1}^T  r_t
\label{reward}
\end{equation}
In AVSR task, we summarize the MDP tuple as:
\begin{itemize}
    \item State $\mathcal{S}$ should contain the comprehensive information or learned patterns for decision-making. Therefore, we denoted $\mathcal{S}$ as a combination of $F_i$, $F_v$, and $I_a$ defined in Section~\ref{3-1}, as they are related to predict current token.
    \item Action $\mathcal{A}$ aims to interact with the environment and update the $\mathcal{S}$. In this work, $\mathcal{A}$ is defined as a probability distribution $P_a$ for the current predicted token.
    \item Reward $\mathcal{R}$ is an instant feedback signal to evaluate the performance of $\mathcal{A}$. we define a token-level reward function for each hypothesis $Y$ as follows:
    \begin{equation}
    \small
    \begin{split}
    \mathcal{R}(Y,Y^*) = - \DED(Y||Y^*) &- \lambda_1 \sum_{t=0}^T \DKL(P_a^t||F_i^t) \\&- \lambda_2  \sum_{t=0}^T \DKL(P_a^t||F_v^t)
    \end{split}
    \label{eq2}
    \end{equation}
    where $\DED(\cdot||\cdot)$ denotes the edit distance between two sequence, and $Y^*$ denotes the ground-truth sequence. It is noted that such distance is directly related to WER. The $\DKL(\cdot||\cdot)$ denotes the KL-divergence between two distributions, which are used to constrain the policy network to explore in trust region~\cite{schulman2015trust}. $\lambda_1$ and $\lambda_2$ are the weights to balance them.
    \item Transition dynamics $\mathcal{T}(s_{t+1}|s_t,a_t)$ denotes that the predicted token $a_t$ will influence next state $s_{t+1}$, since the decoding of AVSR is auto-regressive generation process.
\end{itemize}
In order to maximize the cumulative reward $\mathcal{R}$, the training objective of policy network is defined as:
\begin{equation}
\begin{split}
\mathcal{L}_\theta(\left \langle A,V \right \rangle,Y^*) &= - \mathbb{E}[\mathcal{R}(Y,Y^*)] \\&= \sum_Y \mathcal{P}(Y|\left \langle A,V \right \rangle, \theta) \mathcal{R}(Y,Y^*) 
\end{split}
\label{eq3}
\end{equation}
where $\theta$ denotes the neural network, $\mathcal{P}(Y|\left \langle A,V \right \rangle, \theta)$ is the probability of hypothesis $Y$ determined by input $\left \langle A,V \right \rangle$ and $\theta$. The reward function $\mathcal{R}(Y,Y^*)$ is defined in E.q~(\ref{eq2}).\par
Since $\sum_Y \mathcal{P}(Y|\left \langle A,V \right \rangle, \theta)$ involves a summation over all possible sequences, we employ the REINFORCEMENT algorithm~\cite{williams1992simple} to approximate the expected $\mathcal{R}$ and calculate the gradient $\nabla_\theta \mathcal{L}_\theta$:
\begin{equation}
\small
\nabla_\theta \mathcal{L}_\theta \! \!= \! \!- \mathbb{E}_{Y^n\sim \mathcal{P}(Y^n|\left \langle A,\!V \right \rangle,\theta)} [\mathcal{R}(Y^n\!,Y^*\!)\nabla_\theta log \mathcal{P}(Y^n| \! \left \langle A,\!V \right \rangle\!,\theta)] 
\label{eq3}
\end{equation}
\normalsize
Where $Y^n$ is the sampling hypothesis drawn from the current model distribution. Different from other sampling methods, we directly utilize the beam search algorithm during decoding to select the $N$-best hypothesis, which indicates the number of sampling hypotheses is equal to the beam size $N$. Furthermore, we introduce the baseline to normalize the reward as follows:
\begin{equation}
\small
\nabla_\theta\mathcal{L}_\theta \! = \! -\frac{1}{N} \! \! \sum_{Y^n\in \text{Beam}}^N \! \! \! \nabla_\theta log \mathcal{P}(Y^n|\left \langle A,\!V \right \rangle,\theta) \ [\ \mathcal{R}(Y^n,Y^*)-\Bar{\mathcal{R}} \ ] 
\label{eq5}
\end{equation}
\normalsize
where $\Bar{\mathcal{R}}$ is the baseline defined as the average of reward of all hypotheses in a beam set. Subtracting $\Bar{\mathcal{R}}$ does not influence the gradient, but importantly, it can reduce the variance of the gradient estimation, thus stabilizing the training process. To simplify the calculation, we assume that the probability mass is concentrated on the $N$-best list only. Consequently, the loss function can be approximated as: 
\begin{equation}
\small
\mathcal{L} \approx - \!\!\! \sum_{Y^n\in \text{Beam}}^N \!\! log\hat{\mathcal{P}} (Y^n|\left \langle A,\!V \right \rangle,\theta) \ [\ \mathcal{R}(Y^n,Y^*)-\Bar{\mathcal{R}} \ ] 
\label{eq6}
\end{equation}
where $\hat{\mathcal{P}}(Y^n|\left \langle A,\!V \right \rangle,\theta) = \frac{\mathcal{P}(Y^n|\left \langle A,\!V \right \rangle,\theta)}{\sum_{Y^n\in \text{Beam}} \mathcal{P}(Y^n|\left \langle A,\!V \right \rangle,\theta)}$ represents the re-normalized distribution over the N-best hypotheses. Accordingly, in one Beam set, those hypotheses with a higher reward than average are encouraged to be selected by increasing their possibilities. Conversely, the hypothesis that obtains a lower reward will be suppressed. By minimizing the criterion of E.q~(\ref{eq6}), the current model intends to pursue higher reward by effective exploration in a beam set. 
\subsection{Training Schedule of MSRL}
\label{3-3}
The training process contains two stages as shown in Algorithm~\ref{algo1}. We first use typical cross-entropy criterion to train the randomly initialized decoder that is shown from step 1 to step 4. The best model is selected by a valid set for subsequent sampling. Then the RL optimization is applied to integrate the visual modality-specific representations according to the reward function in step 4. Considering the continuity of utterance, we adopt an online training manner and the gradient is calculated after the completion of the beam search. Consequently, to achieve higher reward, the policy network will be updated to the direction which optimizes the posterior metric.

\begin{algorithm}[h!]
  \caption{Pseudocode for MSRL Training}
  \begin{algorithmic}[1]
    \Require
      The paired audio $A$, video $V$, and corresponding ground-truth sequence $Y^*=(y_1^*,y_1^*,...,y_T^*)$. 
    \State Initialize the pre-trained parameters for AV encoder $\theta_{av}$ and vision model $\theta_v$.  
    \State Initialize the random parameters for Transformer decoder $\theta_d$, MLP block $\theta_m$, and RL policy network $\theta_p$.
    \State \textbf{while} TRUE \textbf{do} \par
    Freeze the parameters of encoder $\theta_{av}$ \par
    Obtain the hidden feature $h_{av}\!=\theta_{av}(A,V)$ \par
    Train the decoder using cross-entropy loss $\mathcal{L}_{ce}$: \par
    \vspace{-0.5cm}
    \begin{equation}
        \mathcal{L}_{ce} = \sum_{t=1}^T - \log  \mathcal{P}_{\theta_d}(y_t^* |\ y_{t-1}^*, ... ,y_1^*,h_{av})
        \vspace{-0.1cm}
    \end{equation} 
    \textbf{end while} \par
    \State \textbf{while} TRUE \textbf{do} \par
    \textbf{for} hypothesis $Y^n$ in $N$-best list \textbf{do} \par
     \hspace{0.6cm}Freeze the encoder $\theta_{av}$ and vision model $\theta_{v}$ \par
     \hspace{0.6cm}\textbf{for} t in 1,2,..., T \textbf{do}\par     
     
     \hspace{1.2cm} Obtain representations $F_i$ and $F_v$: \par
     \hspace{1.2cm} $F_i$ = $\theta_{d}(h_{av})$\par
     \hspace{1.2cm} $F_v$ = $\theta_{v}(V)$\par
     \hspace{1.2cm} Compute current action probability: \par
     \hspace{1.2cm} $P_a^t = \theta_p(F_i^t,F_v^t,\theta_m(A))$ \par
     \hspace{0.6cm}\textbf{end for} \par
     \hspace{0.6cm}Compute probability $\mathcal{P}(Y^n)= \prod_{t=1}^T P_a^t$\par
     \vspace{0.12cm}
     \hspace{0.6cm}Determine accumulative reward $\mathcal{R}(Y^n,Y^*)$ \par
     \textbf{end for} \par
    Train the policy network using E.q (\ref{eq6})\par
    \hspace{-0.6cm}\textbf{end while} \par

    
  \end{algorithmic}
\label{algo1}
\end{algorithm}

\section{Experiment Setting}
\subsection{Database}
We conduct the experiments on LRS3~\cite{afouras2018lrs3}, which is the largest publicly available dataset for audio-visual speech recognition task. It includes face tracks from over 400 hours of TED and TEDx videos from more than 5,000 speakers, along with the corresponding subtitles and word alignment boundaries. The original training set is divided into 2 partitions: pretrain (403 hours) and trainval (30 hours), which are both from the same sources with test set (1452 utterances, 1 hour). In this paper, we randomly choose 1,200 utterances (1 hour) as a valid set for hyper-parameter tuning and best model selection. \par
For the noisy test set, we follow the categories and mixing strategy from prior work~\cite{shi2022robust}. The seen noises contains categories of ``\emph{babble}", ``\emph{music}" and ``\emph{natural}" that are sampled from MUSAN dataset~\cite{snyder2015musan}, and ``\emph{speech}" noise is sampled from utterances in LRS3. These four categories of noises are seen by both pre-trained models and the training process. For the unseen noises, we select 4 categories of ``\emph{Cafe}", ``\emph{Meeting}", ``\emph{River}", and ``\emph{Resto}" from DEMAND noise set~\cite{thiemann2013diverse} and mix them with test set. The detailed data pre-processing strategy is illustrated in appendix. \par 

\subsection{MSRL Set up}
We develop several MSRL frameworks with different settings, as shown in Table~\ref{table0}. The small transformer block has 768/3072/12 of embedding dimension/feed-forward dimension/attention heads, and the large transformer block increases to 2034/4096/16 respectively. Considering the task difficulty of lip reading, the encoder and decoder of vision model adopt large blocks.\par
The labeled data is first used for the pre-trained models~\cite{shi2022learning}, then it is reused for the training of the decoder and RL module. According to labeled data amount, we define it as two modes. The normal-resource contains 433 hours of full training data (pretrain subset and trainval subset), and the low-resource only contains 30 hours of training data (trainval subset). \par

\begin{table}[t]
\resizebox{1.0\columnwidth}{!}{
\begin{tabular}{c|ccc|c}
\toprule
\multirow{2}{*}{ID} & \multirow{2}{*}{\begin{tabular}[c]{@{}c@{}}AV Pre-trained Encoder\\ (\# Enc. blocks )\end{tabular}} & \multirow{2}{*}{\begin{tabular}[c]{@{}c@{}}Decoder\\ (\# Dec. blocks )\end{tabular}} & \multirow{2}{*}{\begin{tabular}[c]{@{}c@{}}Vision Pre-trained Model\\ (\# Enc./ Dec. blocks)\end{tabular}} & \multirow{2}{*}{\begin{tabular}[c]{@{}c@{}}Labeled data\\ (hours)\end{tabular}}  \\
                    &                                                                                                  &                                                                                                         &                               \\ \toprule
1   & Small (12)  & Small (6) & Large (24/16) &  30                      \\ 
2   & Small (12)  & Small (6)                                                                                       & Large (24/16)                                                                                           & 433                     \\ 
3   & Large  (24) & Large (9)                                                                                     & Large (24/16)                                                                                           & 30                      \\ 
4   & Large (24)  & Large (9)                                                                      & Large (24/16)                                                                                           & 433                  \\ \bottomrule
\end{tabular}}
\caption{Different settings of MSRL. ``\# Enc." and ``\# Dec." denotes the numbers of encoder and decoder blocks. }
\vspace{-0.2cm}
\label{table0}
\end{table}

\section{Result and Analysis}
In this section, we conduct extensive experiments and answer the following questions: 
\begin{itemize}
    \item What is the effect of modality-specific representations in AVSR task? We display the experimental results to prove that MSRL addresses the research problem in Fig.~\ref{fig1}, and the problematic situation will not happen at low SNR conditions.
    \item What is the effect of the RL integration? We conduct comparative experiments including other integration strategies to show the superiority of RL method. 
    \item How does the MSRL performance against other competitive methods? We carry out a series of experiments in various conditions to compare our method with previously published works. 
    \item How is the generalization of MSRL to unseen noises? We directly test the MSRL on various conditions with unseen noise to demonstrate its ability of generalization. 
\end{itemize}

\subsection{Effect of Modality-specific Representations}
In this part, we first quantitatively analyze the effect that MSRL utilizes the visual modality representations. To this end, we construct three baseline systems that leverage different representations. \textbf{\emph{Audio-only}} baseline only consumes audio modality as input to generate target sequence. \textbf{\emph{Visual-only}} baseline is trained as a lip-reading task that consumes visual modality as input. \textbf{\emph{Modality-invariant}} baseline is trained as a vanilla AVSR task that consumes both audio and visual modality as input. Considering the intensity, the \emph{babble} noise is selected to simulate the noisy condition with different SNR levels. The WER results of MSRL and three baselines with different Transformer blocks and resource modes are shown in Table~\ref{table1}. \par 

\begin{table}[t]
\centering
\resizebox{1.0\columnwidth}{!}{
\begin{tabular}{c|m{0.76cm}<{\centering}|cccccc|c}
\toprule
\multirow{2}{*}{Method} & \multirow{2}{*}{Block} & \multicolumn{6}{c|}{\emph{Babble} Noise, SNR=} & Clean \\ 
                        &        &-15 & -10    & -5  & 0  & 5  & avg &$\infty$                    \\\midrule
\multicolumn{9}{c}{\cellcolor[HTML]{E0E0E0}Normal-resource (433 hours of labeled data)}  \vspace{0.1cm}   \\                   
\emph{Audio-only}              & Small  & 99.1 &   98.1  &  82.7  & 32.6 & 11.9 & 64.9 & 2.53                    \\ 
\emph{Audio-only}              & Large  & 98.6 & 97.4  &  75.8  & 24.6 & 9.01 & 61.1 & 1.95        \\ 
\emph{Visual-only}             & Large  &\multicolumn{6}{c|}{\cellcolor[HTML]{EFEEEC}26.9} & 26.9                     \\ 
\emph{Modality-invariant}      & Small  & 55.6 & 38.0 & 19.1  & 7.24  & 4.02 & 24.8 & 1.84     \\ 
\emph{Modality-invariant}       & Large  & 43.4 & 30.3  & 13.5  & 4.90& 2.50 & 18.9 &1.45                     \\  \midrule 
MSRL                    & Small  & 26.1 & 24.7 & 14.8  & 5.92  & 3.19 & 14.9 & 1.44              \\ 
MSRL                    & Large  & 25.5 & 22.3 & 11.3 & 4.51 & 2.31 & 13.2 & 1.33    \\    \toprule
\multicolumn{9}{c}{\cellcolor[HTML]{E0E0E0}Low-resource (30 hours of labeled data)}  \vspace{0.1cm}     \\   
\emph{Audio-only}            & Small  & * &  *  &   84.2   &  36.1    & 13.9 &66.8 & 4.69                      \\ 
\emph{Audio-only}            & Large  & * &   98.0   &   77.0  &   25.9 &  14.2 & 63.0 &3.51                    \\ 
\emph{Visual-only}        & Large  & \multicolumn{6}{c|}{\cellcolor[HTML]{EFEEEC} 27.8} & 27.8                     \\ 
\emph{Modality-invariant}       & Small & 53.0  & 39.5  &   21.4  & 10.2 & 5.92 & 26.0 & 4.10                    \\ 
\emph{Modality-invariant}       & Large & 44.8  & 32.3 & 16.4 & 7.37     & 4.87  & 21.1 & 3.27                   \\  \midrule 
MSRL                    & Small & 27.4 & 25.8 &  16.7   & 7.24    & 5.20       & 16.5 &   3.38                  \\ 
MSRL                    & Large & 26.5 & 24.9 &   13.0  & 6.36     & 3.97  & 14.9 & 2.82                \\ \bottomrule
\end{tabular}}
\caption{The WER (\%) results in \emph{babble} noise and clean conditions.``avg" denotes the average performance across all SNR. ``*" denotes the input modality can not be recognized.}
\label{table1}
\end{table}

\begin{table}[t]
\centering
\resizebox{1.0\columnwidth}{!}{
\begin{tabular}{c|cccccc|c}
\toprule
\multirow{2}{*}{Method}  & \multicolumn{6}{c|}{\emph{Babble} Noise, SNR=} & Clean \\ 
                         &   -15 & -10    & -5  & 0  & 5  & avg &$\infty$                    \\\midrule
\multicolumn{8}{c}{\cellcolor[HTML]{E0E0E0}Normal-resource (433 hours of labeled data) \& Large Transformer block}  \vspace{0.1cm}   \\   
\emph{Modality-invariant}       & 43.4 & 30.3  & 13.5  & 4.90& 2.50 & 18.9 &1.45                     \\ \midrule   
\emph{Early fusion}             & 38.2 & 25.8& 12.6 & 5.07 & 2.96 & 16.9 &1.58                   \\ 
\emph{Late fusion}              & 36.7 & 26.2  & 12.2 & 4.78 & 2.70 & 16.5 & 1.68      \\ 
\emph{Model ensemble}           & 31.6 & 23.4 & 11.8 & 5.36 & 3.15 & 15.1&2.26\\      \midrule               

MSRL                    & \textbf{25.5} & \textbf{22.3} & \textbf{11.3} & \textbf{4.51} & \textbf{2.31} & \textbf{13.2} & \textbf{1.33}     \\    \toprule
\multicolumn{8}{c}{\cellcolor[HTML]{E0E0E0}Low-resource (30 hours of labeled data) \& Large Transformer block}  \vspace{0.1cm}   \\   
\emph{Modality-invariant}       & 44.8 & 32.3  & 16.4 & 7.37 & 4.87 & 21.1 & 3.27                     \\ \midrule   
\emph{Early fusion}             & 40.1 & 25.6 & 15.5 & 7.40 & 5.01 & 18.7 & 3.26               \\ 
\emph{Late fusion}              & 38.4 & 25.9 & 13.3 &6.40 & 4.19& 17.6& 3.31             \\ 
\emph{Model ensemble}             &33.7 & 25.4 & 13.6  & 6.77& 4.18 & 16.7 & 3.35    \\      \midrule               
MSRL                    & \textbf{26.5} & \textbf{24.9} & \textbf{13.0} & \textbf{6.36} & \textbf{3.97} & \textbf{14.9} & \textbf{2.82}     \\    \bottomrule
\end{tabular}}
\caption{The WER (\%) results of MSRL and other integration methods in \emph{babble} noise and clean conditions. Best results are in bold.}
\label{table2}
\vspace{-0.2cm}
\end{table}

\begin{figure}[h]
  \centering
  \includegraphics[width=0.45\textwidth]{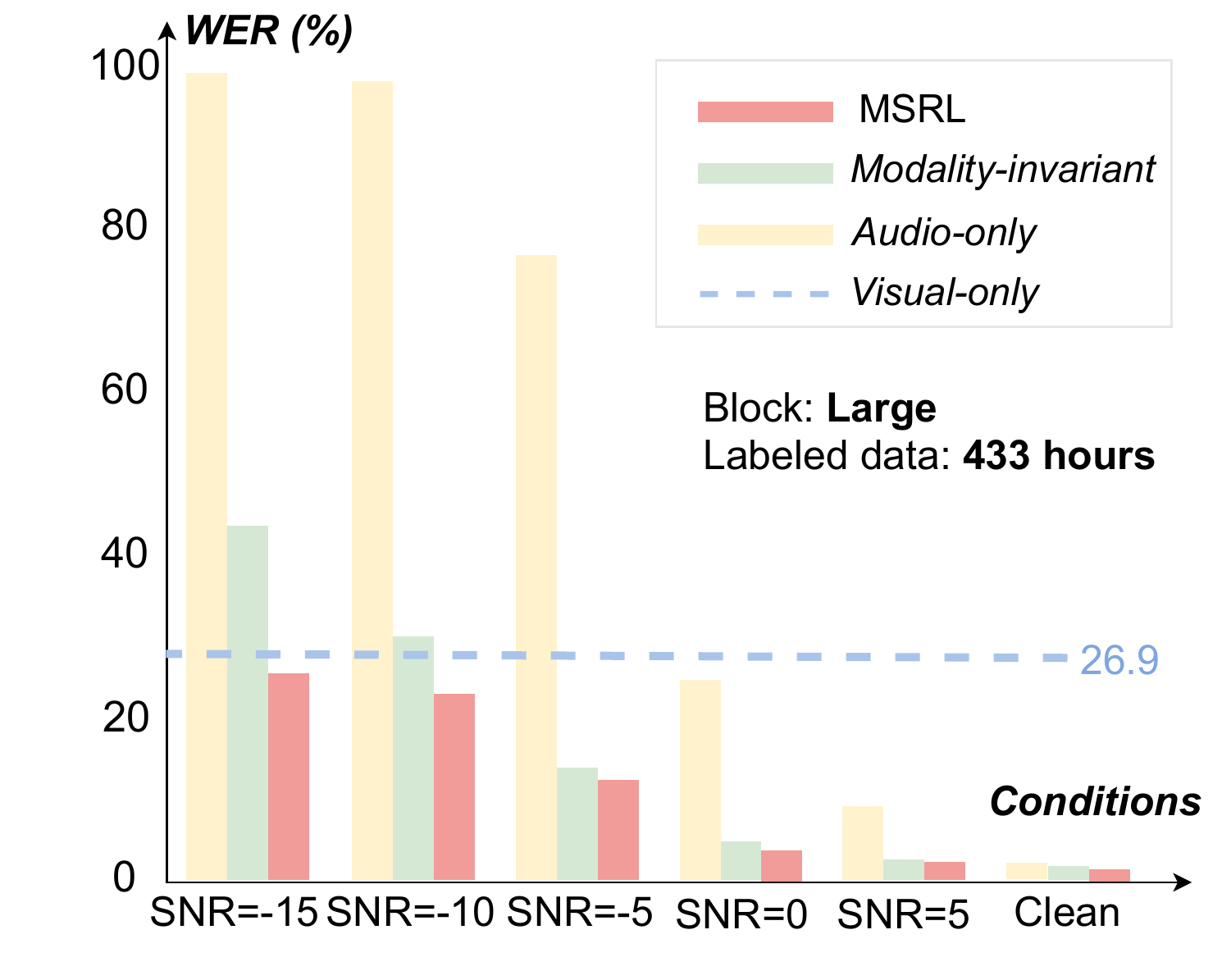}
  \caption{The visualization WER(\%) results of \emph{Audio-only} baseline, \emph{Modality-invariant} baseline, and proposed MSRL with large block and 433 hours of labeled data.}
  \label{fig3}
  \vspace{-0.1cm}
\end{figure}

We observe that except \emph{Visual-only} baseline, the performance of other methods degrades obviously with the decrease of SNR. When SNR is lower than -5, the performance of \emph{Modality-invariant} baseline is even worse than \emph{Visual-only} baseline. However, such a problematic situation does not happen in MSRL, since the visual modality-specific representations have been increasingly effective if audio quality becomes hard to recognize. Furthermore, we observe that MSRL system achieves up to 17.6\% relative WER reduction than \emph{Modality-invariant} baseline in clean conditions. It is out of intuition because the visual modality-specific representations are usually considered trivial when audio quality is high. We reason that 1) visual modality-specific representations add the diversity of information, and it might be helpful when some ambiguous acoustic pronunciations have similar probabilities. 2) The training objective in E.q.(\ref{eq6}) is related to WER, thus ameliorating the mismatch problem between training and testing modes. In general, the proposed MSRL system not only guarantees the lower-bound performance in noisy conditions but also improves the upper-bound performance in clean conditions. \par

In order to visualize the effect of modality-specific representations, we draw the histogram of \emph{Audio-only} baseline, \emph{Modality-invariant} baseline, and MSRL system in Fig.~\ref{fig3}. The \emph{Visual-only} baseline is shown as the blue dashed line as the WER keeps invariant (26.9\%) in all conditions. It is noted that Fig.~\ref{fig3} roughly reproduces the situation in Fig.~\ref{fig1}. The \emph{Modality-invariant} baseline loses its effectiveness in low SNR setting, while MSRL system performs similarly to the oracle line in all conditions.\par
\begin{figure}[h]
  \centering
  \includegraphics[width=0.45\textwidth]{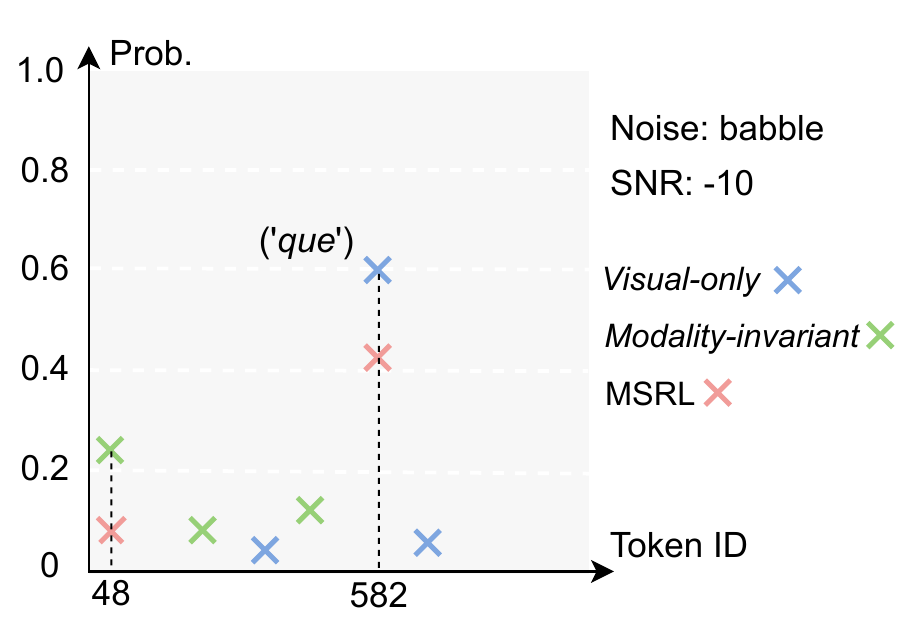}
  \caption{Case study of a divergent decoding step, where the ground-truth token is '\emph{que}' (ID=582). The probabilities higher than 0.05 are displayed. }
  \label{fig4}
  \vspace{-0.1cm}
\end{figure}

We also conduct a case study to observe how the visual modality-specific representations help the MSRL system. To this end, we sample a divergent step in decoding, where the \emph{Modality-invariant} baseline predicts a wrong token but MSRL predict the correct one. As shown in Fig~\ref{fig4}, three probability distributions are drawn from \emph{Modality-invariant} baseline, \emph{Visual-only} baseline, and MSRL system. The x-axis is the vocabulary size and each value denotes a BPE~\cite{sennrich2015neural} token. The y-axis denotes the probability of the corresponding token. For better visualization, the improbable tokens (probability$\textless$ 0.05) are not included in the figure. It is observed that \emph{Modality-invariant} baseline predict a wrong token (ID=48) in this decoding step, but with help of visual modality-specific representations (\emph{i.e.}, \emph{Visual-only} baseline), the MSRL predicts the correct token '\emph{que}' (ID=582).  

\begin{table*}[]
\centering
\resizebox{2.0\columnwidth}{!}{
\begin{tabular}{c|c|ccccc|ccccc|ccccc|ccccc|c}

\toprule
\multirow{2}{*}{Method}& \multirow{2}{*}{Hr} & \multicolumn{5}{c|}{\emph{Babble}, SNR=} & \multicolumn{5}{c|}{\emph{Natural}, SNR=} & \multicolumn{5}{c|}{\emph{Music}, SNR=} & \multicolumn{5}{c|}{\emph{Speech}, SNR=} & Clean \\ 
                       & & -10 & -5 & 0 & 5 &  avg & -10 & -5 & 0 & 5 & avg & -10 & -5 & 0 & 5 & avg & -10 & -5 & 0 & 5 & avg &$\infty$  \\ \midrule
RNN-T        & 34K  & - & - & - & - & - & - & - & - & - & - & - & - & - & - & - & - & - & - & - & - &4.5   \\
TM-seq2seq   & 595  & - & - &   42.5   & - & - & - & - & - & - & - & - & - & - & - & - & - & - & - & - & - &  7.2 \\
AE-MSR       & 1.4K & 38.6 & 31.1 & 25.5 & 24.3 & 29.9 & - & - & - & - & - & - & - & - & - & - & - & - & - & - & - & 6.8  \\
AV-HuBERT    & 30  & 35.1 & 18.4 & 8.3 & 4.9 & 16.7      &   11.6 & 6.5 & 4.6 & 4.0 & 6.7  &  12.4 & 7.4 &  4.7  & 4.1 &   7.2   &     11.5 & 6.8 & 5.0 & 4.2& 6.9& 3.3 \\
AV-HuBERT    & 433  & 34.9 & 16.6 & 5.8 & 2.6 &  15.0    & 9.4 &  4.3  &  2.4 & 1.9 & 4.5  &  10.9 & 4.6 & 2.6 & 1.8 &   5.0   &    11.4 & 4.6 & 2.9 & 2.2 & 5.3 & 1.4  \\
\midrule
MSRL (ours)         & 30 &    24.9   &  13.0 &  6.4  &  4.1  &   12.1    & 9.8 & 5.6 &  3.7 & 3.4  &   5.6   &  10.8  &  6.5 & 4.0 & 3.3 &  6.2    & 8.6 &5.5    &4.0 &3.5 & 5.4 & 2.8   \\
MSRL (ours)         & 433 &     22.4  & 11.3     &   4.5   &  2.3    &   10.1    & 8.0 & 4.1 &   2.3   & 1.6     &  4.0    & 8.9 &   4.4   &   2.4    &  1.7     &    4.4  & 7.2&  3.4  & 2.3 & 1.8 &  3.7 & 1.3   \\ \bottomrule
\end{tabular}}
\caption{The WER (\%) results of MSRL and prior works on LRS3 dataset.``\emph{Hr}" denotes the the amount of labeled audio-visual speech data used in each system. ``\emph{Babble}", ``\emph{Natural}", and ``\emph{Music}" are the different types of noise from MUSAN. ``\emph{Speech}" are sampled from other utterances in LRS3.}
\label{table3}
\end{table*}
\begin{table*}[]
\centering
\resizebox{2.0\columnwidth}{!}{
\begin{tabular}{c|ccccc|ccccc|ccccc|ccccc|ccccc}
\toprule
\multirow{2}{*}{Method} & \multicolumn{5}{c|}{\emph{Cafe}, SNR=} & \multicolumn{5}{c|}{\emph{Meeting}, SNR=} & \multicolumn{5}{c|}{\emph{River}, SNR=} & \multicolumn{5}{c|}{\emph{Resto}, SNR=} & Clean  \\ 
                        & -10 & -5 & 0 & 5 &  avg & -10 & -5 & 0 & 5 & avg & -10 & -5 & 0 & 5 & avg & -10& -5 & 0 & 5 & avg &$\infty$   \\ \midrule
\multicolumn{22}{c}{\cellcolor[HTML]{E0E0E0}Low-resource (30 hours of labeled data) \& Large Transformer block}  \vspace{0.1cm} \\
AV-HuBERT & 16.4 & 7.5 & 4.7 & 4.0 & 8.2&13.6 & 7.3 &4.9 &  4.1 & 7.5   &   23.6   & 11.0  & 5.9 & 4.4& 11.2 &36.8&19.9& 8.3&5.1 &17.5 &3.3\\             
MSRL (ours)      & 13.0 & 6.1 & 3.9 & 3.1 &6.5 & 11.1 & 6.4 &4.4 & 3.4  & 6.3  &  18.5 & 9.5 &  5.0 &  3.7&9.2 & 24.5 &16.3&7.0 & 4.3 & 13.0& 2.8\\ \midrule
\multicolumn{22}{c}{\cellcolor[HTML]{E0E0E0}Normal-resource (433 hours of labeled data) \& Large Transformer block}  \vspace{0.1cm} \\
AV-HuBERT & 13.1 & 4.8 & 2.6 & 1.9 & 5.6 & 12.4  & 5.4 & 3.0 & 2.2 & 5.8& 21.0 &8.3  &  3.6 & 2.4 &8.8 & 35.9 & 17.4&  5.9 & 2.8 & 15.5 & 1.4  \\
MSRL (ours)      &  11.2  & 4.2 &2.3 & 1.7 & 4.9 & 10.4 & 4.5 &2.6& 1.8 & 4.8 &  17.8 & 7.8 &3.2 &1.9 &7.7 & 23.9 &13.9& 5.1&2.4 & 11.3 &1.3\\ \bottomrule
\end{tabular}}
\caption{The WER (\%) results of MSRL on unseen noises. ``\emph{Cafe}", ``\emph{Meeting}", ``\emph{River}", and ``\emph{Resto}" are the different types of noise from DEMAND.}
\label{table4}
\end{table*}

\subsection{Effect of RL Module}
In this part, we examine the effect of RL module by replacing it using other integration methods, which are \emph{early fusion}, \emph{late fusion}, and \emph{Model ensemble}. Since the pre-trained vision model also has the Transformer-based encoder-decoder pipeline, the \emph{Early fusion} adds the hidden features from the final encoder layer of pre-trained vision model to the corresponding layer of pre-trained AV encoder. The \emph{Late fusion}~\cite{inaguma2019transfer} executes a similar operation but add features at the final layer of decoder. Both early and late fusion strategies are applied in the cross-entropy training (step 3 in Algorithms~\ref{algo1}), where the decoder is trainable to fit the new features. The \emph{Model ensemble} method directly computes the average of probabilities from \emph{Modality-invariant} baseline and \emph{Visual-only} baseline for token prediction in auto-regressive decoding, without any tuning operation. \par
From the WER results of Table~\ref{table2}, in noisy conditions, all integration methods can benefit from visual modality-specific representations compared with \emph{Modality-invariant} baseline. MSRL achieves best performance in all SNR levels. Surprisingly, the untrained \emph{Model ensemble} baseline beats the \emph{Early fusion} and \emph{Late fusion} baselines on average in both normal-resource and low-resource modes. When SNR is -15, except MSRL, three other baselines are not able to avoid the problematic situation that perform worse than \emph{Visual-only} baseline. In clean conditions, however, the visual modality-specific representations might be redundant. We observe that \emph{Model ensemble} baseline overestimates the importance of visual modality-specific representations, thus suffering 55.9\% of performance deterioration (1.45\% $\xrightarrow[]{}$ 2.26\%) in normal-resource mode. \emph{Early fusion} and \emph{Early fusion} can dilute it by tunable parameters, thereby obtaining comparable WER results with \emph{Modality-invariant} baseline. In general, MSRL can reasonably balance the importance of modality-specific and modality-invariant representations, as the policy network always considers acoustic information in auto-regressive decoding.\par 

\subsection{Benchmark against Other Methods}
We then report the WER performance of MSRL in various conditions, as well as comparing it with other competitive methods. Four recent published methods are selected as strong baselines, which are RNN-T~\cite{makino2019recurrent}, TM-seq2seq~\cite{afouras2018conversation}, AE-MSR~\cite{xu2020discriminative}, and AV-HuBERT~\cite{shi2022learning}. Since RNN-T and TM-seq2seq methods focus on clean conditions, and the AE-MSR is only evaluated on \emph{babble} noise, we only report the available results from their respective papers. For AV-HuBERT, the ``\emph{babble}", ``\emph{speech}", and ``clean" columns present the WER results from original paper. The ``\emph{natural}" and ``\emph{music}" columns were reproduced using the official code as they are not available in original paper. The comparison of WER results is shown in Table~\ref{table3}.  \par
In clean conditions, we observe that MSRL achieves 5\% (1.4\% $\xrightarrow[]{}$1.33\%) relative WER reduction over the best baseline of AV-HuBERT in normal-resource mode. In low-resource mode, such superiority increases to 14.5\% (3.3\% $\xrightarrow[]{}$2.82\%). It indicates that visual information is particularly important when training data is limited. Furthermore, the MSRL using 30 hours of labeled data even performs better than RNN-T employs 34k hours of labeled data, which shows better data efficiency.\par
In noisy conditions, MSRL achieves the best performances in all kinds of noises and SNR levels. For the ``\emph{babble}", ``\emph{natural}", ``\emph{music}" and ``\emph{speech}" noises, MSRL respectively surpasses AV-HuBERT baseline by 32.5\%/27.5\%, 11.1\%/15\%, 12.0\%/19.5\% and 30.2\%/21.7\% relatively in normal-resource/low-resource mode. It is noted that the \emph{speech} noise is the utterance drawn from the same source of LRS3 which might confuse the recognizer, while the MSRL can address it well without any separation module.  

\subsection{Generalization on Unseen Noise}
Finally, we evaluate the generalization of MSRL method, as the AVSR model usually encounters unseen noises in practical applications. We test the AV-HuBERT and MSRL models on a customized test set which contains 4 types of unseen noises, and the WER results are shown in Table~\ref{table4}.\par
We observe MSRL system has better generalization on all 4 kinds of noises. The visual modality-specific representations are still effective as they are unaffected by the domain shift of audio modality. Consequently, MSRL respectively surpasses the AV-HuBERT baseline by 20.7\%/12.5\%, 16.4\%/17.2\%, 17.9\%/12.5\% and 25.7\%/27.1\% relatively in low-resource/normal-resource mode. Furthermore, we notice that models show distinct adaptability to different unseen noises. Since the ``\emph{cafe}" and ``\emph{meeting}" noises mainly consist of human voice, both AV-HuBERT and MSRL adapt them well and achieve comparable WER results with seen ``\emph{speech}" noise. However, the WER performance degrades obviously on ``\emph{river}" and ``\emph{resto}" noises, as there is no similar seen noise during training process. 
\section{Conclusion}
In this paper, we propose a reinforcement learning-based method MSRL to leverage the modality-specific representations into AVSR task. MSRL employs a pre-trained vision model to provide the visual modality-specific and a policy network to explore the optimal integrated strategy in auto-regressive decoding process. We design the experiments to examine the effects of both visual modality-specific representations and RL integration module. WER results demonstrate that MSRL achieves state-of-the-art performance on LRS3 dataset in clean and noisy conditions, as well as showing better generalization on unseen noises.\par

\section{Acknowledgement}
This research is supported by the National Research Foundation, Singapore
under its AI Singapore Programme (AISG Award No: AISG2-PhD-2021-01-002).

\bibliography{aaai22}

\end{document}